**A short remark on negative energy densities and quantum inequalities.**

Dan Solomon
Rauland-Borg Corporation
3450 W. Oakton
Skokie, IL 60077  USA

Email: **dan.solomon@rauland.com**
(Jan. 19,  2009)

**Abstract**

In quantum field theory it is generally known that the energy density may be negative at a given point in spacetime.  A number of papers have shown that there is a restriction on this energy density which is called a quantum inequality (QI).  A QI is the lower bound to the "weighted average" of the energy density at a given point integrated over a time dependent sampling function.  In this paper we give an example of a sampling function for which there is no QI.

## 1. Introduction

In quantum theory it is widely known that that energy density may be unboundedly negative at a given point in space-time. However it has been shown that there are restrictions on the negative energy density which are referred to as the quantum inequalities (QI) [1]. The QI's have the following form: Let $\langle\Omega|\hat{T}_{00}(\mathbf{x},t)|\Omega\rangle$ be the energy density of the normalized state vector $|\Omega\rangle$ where $\hat{T}_{00}(\mathbf{x},t)$ is the Heisenberg picture energy density operator. Let $s(t)$ be a peaked non-negative sampling function whose time integral is unity. The QI takes the form,

$$T_{00,Ave} \equiv \int_{-\infty}^{+\infty} \langle\Omega|\hat{T}_{00}(\mathbf{x},t)|\Omega\rangle s(t) dt \geq F(s(t)) \qquad (1.1)$$

That is, there is a lower bound to the quantity $T_{00,Ave}$ which is defined by the above equation. This lower bound is given by the quantity $F(s(t))$ which is dependent on the sampling function $s(t)$. Ford and Roman [1][2] have considered the specific case where the sampling function is given by $s(t) = t_0 / \left[\pi(t^2 + t_0^2)\right]$. Now what happens if we pick a different sampling function? Should we assume that a QI exists for all sampling functions or is there something unique about the sampling function specified in the last sentence?

This problem was originally addressed for a zero mass scalar field theory in 1-1D spacetime by E.E. Flanagan [3]. Fewster and Eveson [4] extended this work to 4 dimensions and non-zero mass field. It will be shown in the following discussion that for a zero mass scalar field in four dimensional space-time we can find a sampling function for which there is no lower bound to $T_{00,Ave}$.

## 2. Calculating the energy-density.

We will consider scalar field theory with zero mass in four dimensional spacetime. We will start by examining the quantity $\langle\Omega|\hat{T}_{00}(0,t)|\Omega\rangle$ which is the expectation value of the energy density at the origin $\mathbf{x}=0$. Referring to Eq. 7 of Ref [1] it can be shown that for a zero mass scalar field,

$$\langle\Omega|\hat{T}_{00}(0,t)|\Omega\rangle = \frac{\text{Re}}{2V}\sum_{\mathbf{k},\mathbf{q}}\left\{\frac{\omega_k\omega_q + \mathbf{kq}}{\sqrt{\omega_k\omega_q}}\left[\langle\Omega|\hat{a}_\mathbf{k}^\dagger\hat{a}_\mathbf{q}|\Omega\rangle e^{-i(\omega_q-\omega_k)t} + \langle\Omega|\hat{a}_\mathbf{k}\hat{a}_\mathbf{q}|\Omega\rangle e^{-i(\omega_q+\omega_k)t}\right]\right\} \quad (2.1)$$

where $V$ is the integration volume and where $\omega_k = |\mathbf{k}|$. Also $\hat{a}_\mathbf{k}$ and $\hat{a}_\mathbf{k}^\dagger$ are the usual destruction and creation operators, respectively. They satisfy the commutation relationships

$$\left[\hat{a}_\mathbf{k},\hat{a}_\mathbf{q}^\dagger\right] = \delta_{\mathbf{kq}}; \quad \left[\hat{a}_\mathbf{k},\hat{a}_\mathbf{q}\right] = \left[\hat{a}_\mathbf{k}^\dagger,\hat{a}_\mathbf{q}^\dagger\right] = 0 \quad (2.2)$$

The vacuum state is defined by $|0\rangle$ and is destroyed by the destruction operators, i.e., $\hat{a}_\mathbf{k}|0\rangle = 0$.

Let $|\Omega\rangle$ be given by $|\Omega\rangle = \hat{U}|0\rangle$ where $\hat{U}$ is defined by,

$$\hat{U} = e^{\hat{C}} \quad (2.3)$$

and where $\hat{C}$ is,

$$\hat{C} = \sum_\mathbf{k}\frac{f_\mathbf{k}}{2}\left(\hat{a}_\mathbf{k}^\dagger\hat{a}_\mathbf{k}^\dagger - \hat{a}_\mathbf{k}\hat{a}_\mathbf{k}\right) \quad (2.4)$$

In the above expression $f_\mathbf{k}$ is a real valued constant which will be specified later.

Let the sampling function be given by,

$$s(t) = \begin{cases} Ae^{\lambda_1 t} & \text{for } t < 0 \\ Ae^{-\lambda_2 t} & \text{for } t \geq 0 \end{cases} \quad (2.5)$$

where,

$$A = \frac{\lambda_1\lambda_2}{\lambda_1 + \lambda_2} \quad (2.6)$$

The sampling function obeys the relationship $\int_{-\infty}^{+\infty} s(t)dt = 1$. Given the above we want to determine if there is a lower bound to $T_{00,Ave}$ for this sampling function. To determine this proceed as follows.

It is evident that $\hat{C}^\dagger = -\hat{C}$ therefore $\hat{U}^\dagger = \hat{U}^{-1} = e^{-\hat{C}}$. This means that $\hat{U}$ is a unitary operator and satisfies $\hat{U}\hat{U}^\dagger = \hat{U}^\dagger\hat{U} = 1$. Define the quantity,

$$\hat{b}_\mathbf{k} = \hat{U}^\dagger\hat{a}_\mathbf{k}\hat{U} = e^{-\hat{C}}\hat{a}_\mathbf{k}e^{\hat{C}} \quad (2.7)$$

From this we obtain,

$$\hat{b}_{\mathbf{k}}^{\dagger} = \hat{U}^{\dagger}\hat{a}_{\mathbf{k}}^{\dagger}\hat{U} = e^{-\hat{C}}\hat{a}_{\mathbf{k}}^{\dagger}e^{\hat{C}} \tag{2.8}$$

Using the Baker-Campbell-Hausdorff relationships to expand (2.7) we obtain,

$$\hat{b}_{\mathbf{k}} = \hat{a}_{\mathbf{k}} + \left[\hat{a}_{\mathbf{k}},\hat{C}\right] + \frac{1}{2!}\left[\left[\hat{a}_{\mathbf{k}},\hat{C}\right],\hat{C}\right] + \frac{1}{3!}\left[\left[\left[\hat{a}_{\mathbf{k}},\hat{C}\right],\hat{C}\right],\hat{C}\right] + \ldots \tag{2.9}$$

Use (2.2) and (2.4) in the above to obtain,

$$\left[\hat{a}_{\mathbf{k}},\hat{C}\right] = f_{\mathbf{k}}\hat{a}_{\mathbf{k}}^{\dagger} \text{ and } \left[\hat{a}_{\mathbf{k}}^{\dagger},\hat{C}\right] = f_{\mathbf{k}}\hat{a}_{\mathbf{k}} \tag{2.10}$$

Use this in (2.9) to yield,

$$\hat{b}_{\mathbf{k}} = \hat{a}_{\mathbf{k}} + f_{\mathbf{k}}\hat{a}_{\mathbf{k}}^{\dagger} + \frac{f_{\mathbf{k}}^{2}}{2!}\hat{a}_{\mathbf{k}} + \frac{f_{\mathbf{k}}^{3}}{3!}\hat{a}_{\mathbf{k}}^{\dagger} + \ldots = \hat{a}_{\mathbf{k}}\cosh f_{\mathbf{k}} + \hat{a}_{\mathbf{k}}^{\dagger}\sinh f_{\mathbf{k}} \tag{2.11}$$

Also we can show that,

$$\hat{b}_{\mathbf{k}}^{\dagger} = \hat{a}_{\mathbf{k}}^{\dagger}\cosh f_{\mathbf{k}} + \hat{a}_{\mathbf{k}}\sinh f_{\mathbf{k}} \tag{2.12}$$

Use the above results to obtain,

$$\langle\Omega|\hat{a}_{\mathbf{k}}^{\dagger}\hat{a}_{\mathbf{q}}|\Omega\rangle = \langle 0|\hat{U}^{\dagger}\hat{a}_{\mathbf{k}}^{\dagger}\hat{a}_{\mathbf{q}}\hat{U}|0\rangle = \langle 0|\hat{U}^{\dagger}\hat{a}_{\mathbf{k}}^{\dagger}\hat{U}\hat{U}^{\dagger}\hat{a}_{\mathbf{q}}\hat{U}|0\rangle = \langle 0|\hat{b}_{\mathbf{k}}^{\dagger}\hat{b}_{\mathbf{q}}|0\rangle \tag{2.13}$$

Similarly we can also show that,

$$\langle\Omega|\hat{a}_{\mathbf{k}}\hat{a}_{\mathbf{q}}|\Omega\rangle = \langle 0|\hat{b}_{\mathbf{k}}\hat{b}_{\mathbf{q}}|0\rangle \tag{2.14}$$

From (2.11) and (2.12) we obtain,

$$\langle 0|\hat{b}_{\mathbf{k}}^{\dagger}\hat{b}_{\mathbf{q}}|0\rangle = \delta_{\mathbf{kq}}\left(\sinh f_{\mathbf{k}}\right)^{2} \tag{2.15}$$

and,

$$\langle 0|\hat{b}_{\mathbf{k}}\hat{b}_{\mathbf{q}}|0\rangle = \delta_{\mathbf{q}k}\left(\cosh f_{\mathbf{k}}\right)\left(\sinh f_{\mathbf{k}}\right) \tag{2.16}$$

Next, use the above relationships in (2.1) to obtain,

$$\langle\Omega|\hat{T}_{00}(0,t)|\Omega\rangle = \frac{\text{Re}}{2V}\sum_{\mathbf{k}}\left\{2|\mathbf{k}|\left[\left(\sinh f_{\mathbf{k}}\right)^{2} + \left(\cosh f_{\mathbf{k}}\right)\left(\sinh f_{\mathbf{k}}\right)e^{-2i|\mathbf{k}|t}\right]\right\} \tag{2.17}$$

Use,

$$\int_{-\infty}^{+\infty} s(t)e^{-2i|\mathbf{k}|t}dt = A\left[\frac{1}{(\lambda_{1}-2i|\mathbf{k}|)} + \frac{1}{(\lambda_{2}+2i|\mathbf{k}|)}\right] = A\left[\frac{\lambda_{1}+2i|\mathbf{k}|}{(\lambda_{1}^{2}+4|\mathbf{k}|^{2})} + \frac{\lambda_{2}-2i|\mathbf{k}|}{(\lambda_{2}^{2}+4|\mathbf{k}|^{2})}\right] \tag{2.18}$$

and (2.17) in (1.1) to obtain,

$$T_{00,Ave} = \frac{\text{Re}}{2V}\sum_{\mathbf{k}}\left\{2|\mathbf{k}|(\cosh f_{\mathbf{k}})(\sinh f_{\mathbf{k}})\left[\frac{(\sinh f_{\mathbf{k}})}{(\cosh f_{\mathbf{k}})} + A\left(\frac{\lambda_1 + 2i|\mathbf{k}|}{\left(\lambda_1^2 + 4|\mathbf{k}|^2\right)} + \frac{\lambda_2 - 2i|\mathbf{k}|}{\left(\lambda_2^2 + 4|\mathbf{k}|^2\right)}\right)\right]\right\}$$

(2.19)

This yields,

$$T_{00,Ave} = \frac{1}{V}\sum_{\mathbf{k}}\left\{|\mathbf{k}|(\cosh f_{\mathbf{k}})(\sinh f_{\mathbf{k}})\left[\frac{(\sinh f_{\mathbf{k}})}{(\cosh f_{\mathbf{k}})} + A\left(\frac{\lambda_1}{\left(\lambda_1^2 + 4|\mathbf{k}|^2\right)} + \frac{\lambda_2}{\left(\lambda_2^2 + 4|\mathbf{k}|^2\right)}\right)\right]\right\}$$

(2.20)

Next assume that $f_{\mathbf{k}}$ is a function of $|\mathbf{k}|$ and let $V \to \infty$ and make the substitution $\sum_{\mathbf{k}} G(|\mathbf{k}|) \to \frac{V}{(2\pi)^3}\int_0^{+\infty} 4\pi|\mathbf{k}|^2 G(|\mathbf{k}|) d|\mathbf{k}|$ to obtain,

$$T_{00,Ave} = \int_0^{+\infty}\frac{d|\mathbf{k}|}{2\pi^2}\left\{|\mathbf{k}|^3(\cosh f_{\mathbf{k}})(\sinh f_{\mathbf{k}})\left[\frac{(\sinh f_{\mathbf{k}})}{(\cosh f_{\mathbf{k}})} + A\left(\frac{\lambda_1}{\left(\lambda_1^2 + 4|\mathbf{k}|^2\right)} + \frac{\lambda_2}{\left(\lambda_2^2 + 4|\mathbf{k}|^2\right)}\right)\right]\right\}$$

(2.21)

Let $f_{\mathbf{k}} = -g_{\mathbf{k}}$ where $g_{\mathbf{k}} \geq 0$. Therefore,

$$T_{00,Ave} = \int_0^{+\infty}\frac{d|\mathbf{k}|}{2\pi^2}\left\{|\mathbf{k}|^3(\cosh g_{\mathbf{k}})(\sinh g_{\mathbf{k}})\left[\frac{(\sinh g_{\mathbf{k}})}{(\cosh g_{\mathbf{k}})} - A\left(\frac{\lambda_1}{\left(\lambda_1^2 + 4|\mathbf{k}|^2\right)} + \frac{\lambda_2}{\left(\lambda_2^2 + 4|\mathbf{k}|^2\right)}\right)\right]\right\}$$

(2.22)

Next define the constants $W$ and $\Lambda$ where $\Lambda \gg W$ and $W \gg \lambda_1, \lambda_2$ and $W \gg 1$. Define $g_k$ by,

$$g_{\mathbf{k}} = 0 \text{ for } |\mathbf{k}| > \Lambda \text{ and } W > |\mathbf{k}| \qquad (2.23)$$

and,

$$\frac{\sinh g_{\mathbf{k}}}{\cosh g_{\mathbf{k}}} = \frac{A}{2}\left(\frac{\lambda_1}{\left(\lambda_1^2 + 4|\mathbf{k}|^2\right)} + \frac{\lambda_2}{\left(\lambda_2^2 + 4|\mathbf{k}|^2\right)}\right) \text{ for } \Lambda > |\mathbf{k}| > W \qquad (2.24)$$

Use the above relationships in (2.22) to obtain,

$$T_{00,Ave} = -\int_0^{+\infty} \frac{d|\mathbf{k}|}{2\pi^2} \left\{ |\mathbf{k}|^3 (\cosh g_\mathbf{k})^2 \left[ \left(\frac{A}{2}\right)^2 \left( \frac{\lambda_1}{(\lambda_1^2 + 4|\mathbf{k}|^2)} + \frac{\lambda_2}{(\lambda_2^2 + 4|\mathbf{k}|^2)} \right)^2 \right] \right\} \quad (2.25)$$

Next, use the fact that $W \gg \lambda_1, \lambda_2$ and $W \gg 1$ to show that, in the above integral, we can substitute $(\cosh g_k) \cong 1$ and $(\lambda_1^2 + 4k^2) \cong 4k^2$ and $(\lambda_2^2 + 4k^2) \cong 4k^2$. Use these approximations in the above integral to yield,

$$T_{00,Ave} \cong -\frac{A^2}{8\pi^2} \int_W^\Lambda d|\mathbf{k}| \left\{ |\mathbf{k}|^3 \frac{(\lambda_1 + \lambda_2)^2}{(16|\mathbf{k}|^4)} \right\} = -\frac{A^2 (\lambda_1 + \lambda_2)^2}{128\pi^2} \ln\left(\frac{\Lambda}{W}\right) \quad (2.26)$$

Use (2.6) in the above to obtain,

$$T_{00,Ave} \cong -\frac{(\lambda_1 \lambda_2)^2}{128\pi^2} \ln\left(\frac{\Lambda}{W}\right) \quad (2.27)$$

By making $\Lambda$ arbitrarily large it is evident that there is no lower bound on $T_{00,Ave}$.

### **3. Conclusion**

We have examined the energy density for a zero mass scalar field. We have shown that there is no lower bound to the "weighted average" of the energy density for the sampling function given by (2.5). Therefore a QI does not exist for this particular sampling function.

**References.**